# Scaling relation for the bond length, mass density, and packing order of water ice

Chang Q Sun[1,2,3,5,*], Yongli Huang[2], Xi Zhang[1,3], Zengsheng Ma[2], Yichun Zhou[2], Ji Zhou[4], Weitao Zheng[5,*]

1. School of Electrical and Electronic Engineering, Nanyang Technological University, Singapore 639798
2. Key Laboratory of Low-Dimensional Materials and Application Technologies, and Faculty of Materials, Optoelectronics and Physics, Xiangtan University, Hunan 411105, China
3. College of Materials Science and Engineering, China Jiliang University, Hangzhou 310018, China
4. State Key Laboratory of New Ceramics and Fine Processing, Department of Materials Science and Engineering, Tsinghua University, Beijing 100084, China
5. School of Materials Science, Jilin University, Changchun 130012, China

**The packing order of molecules and the distance between adjacent oxygen atoms ($d_{OO}$) in water and ice are most basic yet puzzling. Here we present a scaling solution for this purpose based only on the mass density $\rho$(gcm$^{-3}$),**

$$\begin{cases} d_{OO} = d_L + d_H = 2.6948\rho^{-1/3} & (1) \\ d_L = 2.5621 \times \left[1 - 0.0055 \times \exp(d_H/0.2428)\right] & (2) \end{cases}$$

**where $d_L$ is the length (Å) of the O:H van der Waals bond and $d_H$ the H-O polar-covalent bond projecting on the O---O line. Validated by the measured proton symmetrization of compressed ice, $d_{OO}$ of water and ice, and $d_{OO}$ expansion at water surface, this solution confirms that the fluctuated, tetrahedrally-coordinated structure is unique for water ice.**

Distance between the adjacent oxygen atoms (i.e. O:H-O bond length) in water ice has yet been certain in the range of 2.70 to 3.0 Å[1-13] and the H-O length varies from 0.97 to 1.001 Å[14]. The packing order of molecules in liquid water varies with the snapshot time scale in measurements[15-21]. The structure of liquid water remained a debating issue in terms of the mono-phase of fluctuated, tetrahedrally-coordinated structure[22,23] and the mixed-phase of low- and high- density fragments with thermal modulation of the fragmental ratios[16,24].

In fact, uncertainties in the packing order and in the O:H-O bond length determine uniquely water-ice's density that is relatively easy to be determined. Therefore, one should be able to resolve the uncertain issues from the certainly known mass density.

The packing of water molecules in water should follow the Ice Rule[25]. Fig 1a illustrates an ideal tetrahedron that contains two equivalent water molecules linked by the O:H-O bonds[25,26]. An oxygen atom hybridizes its sp-orbit to form four directional orbits upon reacting with other less electronegative atoms[27,28]. An oxygen atom catches two electrons from neighboring H atoms to form two intra-atomic H-O polar-covalent bonds and fills up the rest two with its nonbonding electron lone pairs ":" to form the inter-molecular O:H bond through van der Waals (vdW) force.

An oxygen ion always tends to find four neighbors to stabilize but the nonequivalent bond angles[27] and the repulsion between the electron pairs on oxygen[26] frustrate this happening in the liquid phase. Therefore, water structure fluctuates with switching on and off of the O:H bonds.

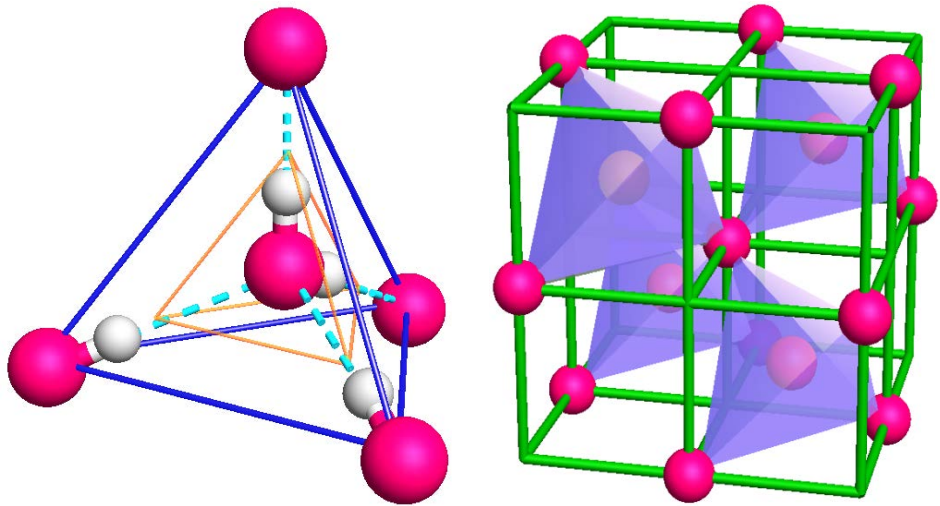

**Fig 1 Packing order of water molecules.** The (left) ideal tetrahedron contains two equivalent H$_2$O molecules connected by O:H-O bonds albeit orientation. The central tetrahedron is Pauling's Ice model[25]. The basic building blocks pack up in a diamond-structure order (right). Therefore, a total of eight H$_2$O molecules occupy this complex cell of eight cubes of a$^3$ volume each. The adjacent oxygen atoms is separated by d$_{OO}$ = √3a/2. The unique packing order and the flexible length determine uniquely the density of water and ice.

Fig 1b shows that four of the eight cubes are occupied by the building blocks and the rest four are empty. Such an ideally diamond order meets the directional specificity of the central oxygen ion. Therefore, the eight cubes of each a$^3$ volume accommodate a total number of eight water molecules. This structure and the O:H-O interaction hold for all phases, from gaseous to ice, unless at extremely high temperature or high pressure[29]. Phase ordering happens if the symmetry or the bond orientation changes[25].

With the known mass of a water molecule consisting 9 pairs of neutrons and protons, M = 9×(1.672621+1.674927)×10$^{-27}$ kg. The known density ρ = M/a$^3$ = 1 (10$^3$kgm$^{-3}$) at 4 °C and the given structure order in Fig (1), gives immediately the density dependence of the $d_{OO}$ in eq (1). The mean value of 2.6948 Å suits only for bulk water at states of statistically stable [15-18].

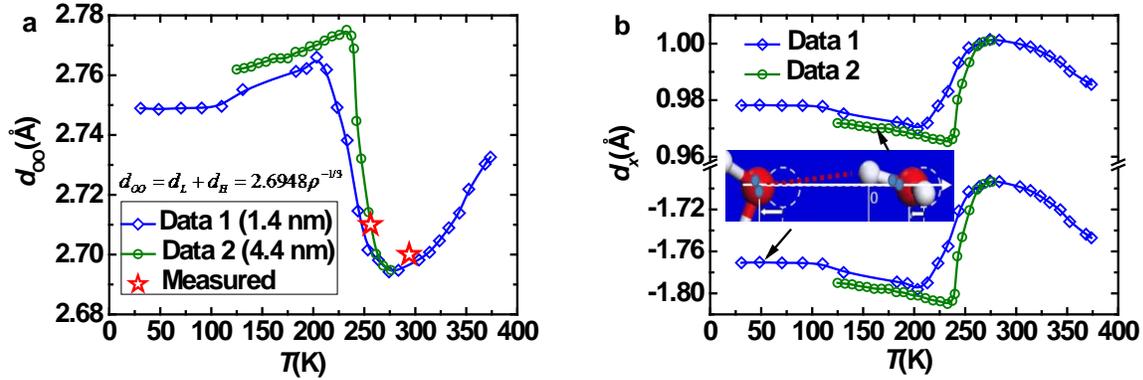

**Fig 2 Density and the specific packing order determine (a) the $d_{OO}$ and (b) the $d_x$ of the O:H-O bond of water ice at cool.** Data 1 corresponds to 1.4 nm[30] and Data 2 to the central of 4.4 nm sized water droplet[31]. Matching the ρ(T)-derived $d_{OO}$ to the direct measurements[1] (2.70 Å at 25 °C and 2.71 Å at -16.8 °C) validates the specific packing order in Fig 1. The $d_x$ corporative relaxation (b) confirms that both O ions displace in the same direction along the O:H-O bond [26]. Inset (b) shows the segmented O:H-O bond with pairs of dots denoting electron pairs on oxygen. H atom is the coordination origin.

Fig 2a shows the $d_{OO}(T)$ resolved from the ρ(T) profiles for the confined water-droplet of different sizes[30,31] using eq (1). The match of the derived $d_{OO}(T)$ to the value of 2.70 Å measured at 25°C and 2.71 Å at -16.8 °C[1] validates that both eq (1) and the packing order in Fig. 1 are essentially true.

Recently, a reproduction of the V(P) profile of compressed ice[32] using molecular dynamics (MD) computation[26] resulted in a decomposition of the V(P) profile into the $d_x(P)$ curves (x = L for O:H and H for H-O bond), see supplementary information (SI) [33]. Consistency between the MD-derived and the measured proton symmetrization, $d_L = d_H = 1.1$ Å of ice under 59~60 GPa[34,35] validates the derived $d_x(P)$ to represent the true situation of $d_L$ and $d_H$ cooperative relaxation.

Plotting the validated $d_L(P)$ against the $d_H(P)$ yields immediately eq (2), which is operating condition (pressure) independent. Using eq (1) and (2), one is able to gain the the $d_L$, the $d_H$, and the $d_{OO}$ with a given density profile. If the derived $d_{OO}$ or $d_H$ match those measurements, then the structure order in Fig 1 and the solution of eqs (1) and (2) are justified true and reliable.

Fig 2b decomposes the $d_{OO}$ into the $d_x$ of water at cool[30,31]. The inset shows the O:H-O hydrogen bond that consists the O:H van der Waals bond and the H-O polar-covalent bond other than either of them alone. Pairs of dots on oxygen atoms are the electron pairs. The decomposed $d_x(T)$ profiles indicates that oxygen atoms dislocate in the same direction but by different amounts with respect to the H atom of coordination origin.

Fig 3 summarizes the $d_L$ and $d_H$ correlation derived using eq (2) from the $\rho(P)$ for ice under compression[32], the $\rho(T)$ for water at cool[30,31]. The $d_H$ of 1.0004 Å at density unity is within the measured values ranging from 0.97 to 1.001 Å[14].

The documented $d_{OO}$ values are often greater[2] than the specified range of 2.6948 and 2.775 Å for water and ice at the atmospheric pressure. Wilson et al[9] uncovered firstly that the surface $d_{OO}$ expands by 5.9% from 2.801 to 2.965 Å at room temperature. Considering the shortest distance of 2.70 Å[1] and the longest 2.965 Å[9], the surface $d_{OO}$ expands by 10% to form the low-density phase in the skin of water. Thus, all reported $d_{OO}$ values as discussed in[1] are correct because of the surface effects. This correlation, eq (2) decomposes the longer $d_{OO} = d_L + d_H$ (>2.82 Å) into the shorter $d_H$ (< 0.95 Å) and the longer $d_L$. Therefore, the $d_H$ contracts for molecular clusters and water surface, following the bond contraction rule of Goldschmidt[36] and Pauling[37].

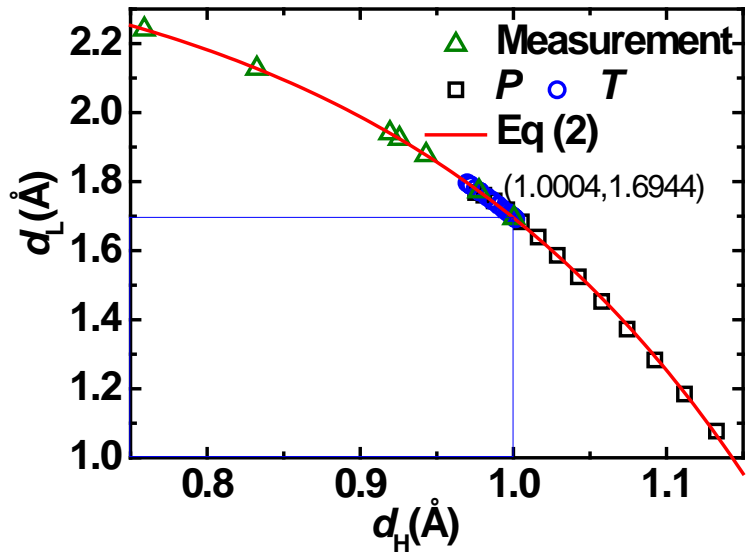

**Fig 3 Universal $d_L$ and $d_H$ relationship of $H_2O$.** Data are derived from the $\rho(P)$[32] and $\rho(T)$[30,31], and direct measurements from liquid and solid[1-8]. The derived $d_H$ = 1.0004 Å at $\rho$ = 1 is within the measured values ranging from 0.97 to 1.001 Å[14]. The $d_H$ shorter than 0.95 Å corresponds to the low-density phase of dimers (2.98 Å), clusters, and surface of water[9,10]

The asymmetric and incorporative relaxation of the $d_x$ is common to water and ice independent of the phase or the testing conditions, which unifies the pressure, temperature, and the size effect on the O:H-O bond length and the incorporative relaxation of the $d_x$. This straightforward yet simple solution to the order-length uncertainties of $H_2O$ has thus been established and justified, which should help in gaining consistent and deeper insight into the unusual behavior of water and ice.

- ASSOCIATED CONTENT

\* Supporting Information
Further information is provided regarding details of the inter-electron-pair repulsion of water and background information as well as nomenclatures regarding basic concepts published previously but not covered in the main text. This material is available free of charge via the Internet at.

- AUTHOR INFORMATION

Corresponding Author
wtzheng@jlu.edu.cn; ecqsun@ntu.edu.sg; CQ is affiliated with honorary appointments at 2, 3, and 5.

Notes

The authors declare no competing financial interest.

- ACKNOWLEDGMENTS

Financial support from the NSF China (Nos.: 21273191, 1033003, 90922025) is gratefully acknowledged.